\documentclass[preprint]{elsarticle}

\newcommand{\beq}{\begin{equation}}
\newcommand{\eeq}{\end{equation}}
\newcommand{\bea}{\begin{eqnarray}}
\newcommand{\eea}{\end{eqnarray}}

\usepackage{graphicx}
\begin{document}
\pagebreak
\title{Tidal waves as yrast states in transitional nuclei.  }
\author{S. Frauendorf, Y. Gu, and J. Sun}
\address{Dept. of Physics, University of Notre Dame, Notre Dame, IN 46556}

\date{\today}

\begin{abstract} The yrast states of transitional nuclei are described as 
quadrupole waves running over the nuclear surface, which we call  tidal waves.
In contrast to a rotor, which generates angular  momentum 
by increasing the angular velocity at approximately constant deformation, 
a tidal wave  generates angular momentum by increasing the deformation
at approximately constant angular velocity. 
The properties of the tidal waves
are calculated by means of the cranking model in a microscopic way. 
The calculated energies and E2 transition probabilities of the yrast
states in the transitional nuclides with $Z$= 44, 46, 48 and
$N=56, ~58,~...,~ 66$ reproduce the experiment in detail. The nonlinear response of 
the nucleonic orbitals results in a strong coupling between shape and single particle 
degrees of freedom.
 
\end{abstract}


\maketitle


Most of the theoretical interpretation of the low-spin structure of nuclei
in the transitional region between spherical and well deformed shape has been
carried out in the framework of phenomenological models based on the Bohr - Hamiltonian\cite{BMII,DCM,Caprio} or using the algebraic IBM concept\cite{IBM,Iachello,castentriangle}.
 Although these studies account  
for  many low-spin aspects of the collective quadrupole degree of freedom
 in an impressive way, 
the connection with the underlying microscopic structure
has remained a challenge. On the other hand, the various versions of 
mean field theory are very successful in predicting 
shell closures, static deformations,
and  delineating the  regions of spherical and deformed shapes 
(for recent reviews cf. e.g. \cite{Moeller,Pearson,Stoitsov}). 
Yet the large amplitude collective dynamics of transitional nuclei
starting from the microscopic mean field remains a challenge to
 nuclear theory. At present there are two mature approaches of
this type:
the combination of mean field potential energy with cranking mass parameters
(see review \cite{BohrHamRev}) and the Generator Coordinate Method (see \cite{GCM1,GCM2}
for recent development).  
Both approaches use the adiabatic approximation that the collective quadrupole 
excitations do not mix with the quasiparticle excitations, 
which is often not justified.   
\begin{figure}[t]
\includegraphics[width=\columnwidth]{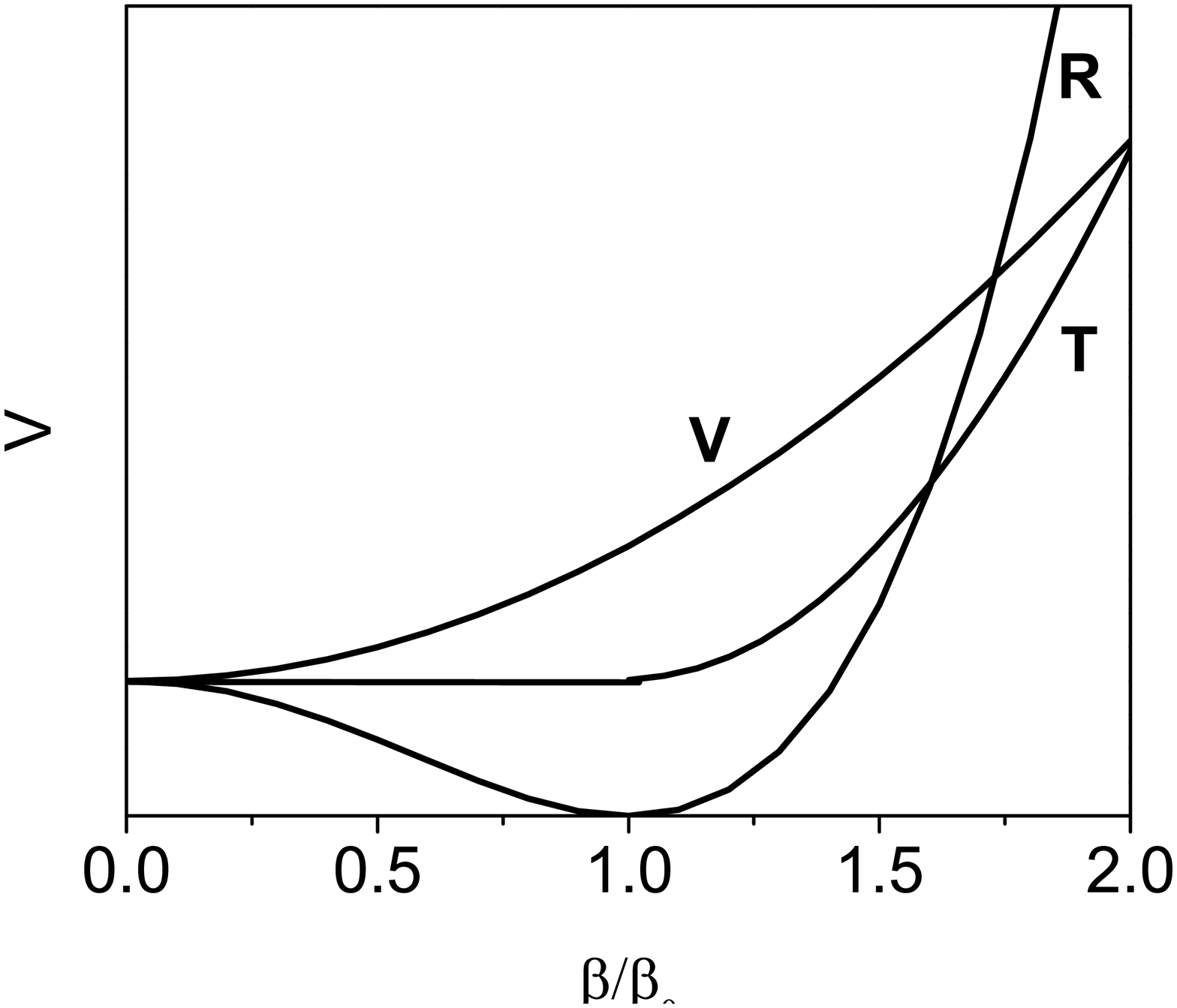}
\caption{Potentials of a vibrator (V), soft rotor (R), and transitional nucleus (T). }
\label{f:potential}       
\end{figure}

In this Letter, we demonstrate that the 
yrast states of transitional nuclei
have the very simple structure of a running surface wave (tidal wave),
which permits us to calculate their properties 
microscopically by means of the cranked mean field theory without resorting to the adiabatic approximation. First we introduce 
the concept of tidal waves for 
the collective quadrupole vibrations of a classical droplet with irrotational
flow, which is discussed in \cite{BMII,DCM}. Then we will extend it to the microscopic 
mean field theory and carry out calculations for a group of transitional nuclei
around $A$=110.   

\begin{figure}[h]
\includegraphics[width=\columnwidth]{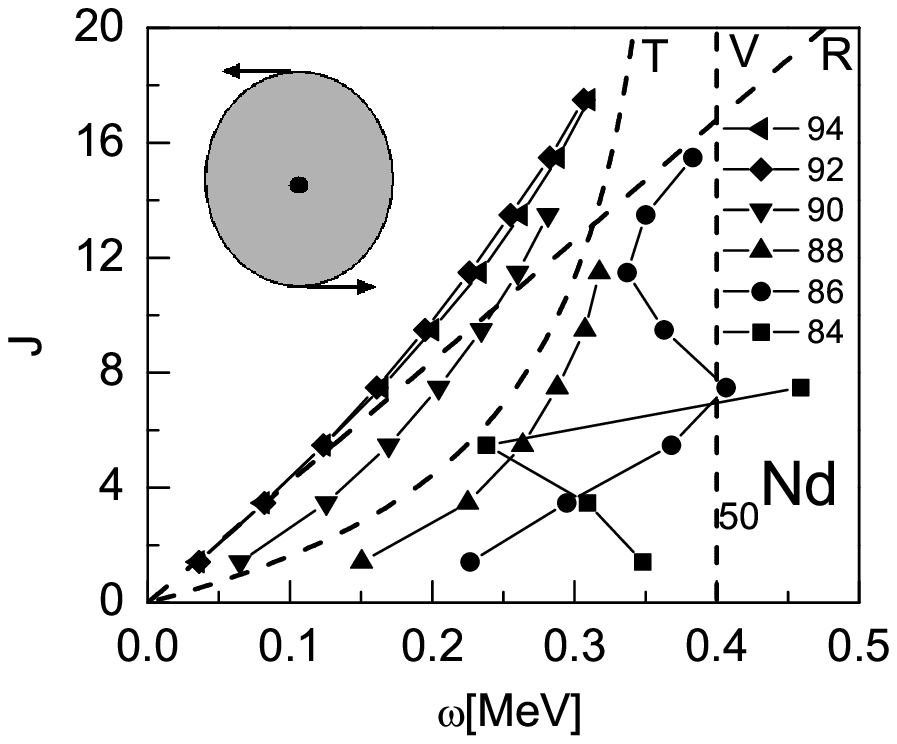}
\vspace*{-1cm}
\includegraphics[width=\columnwidth]{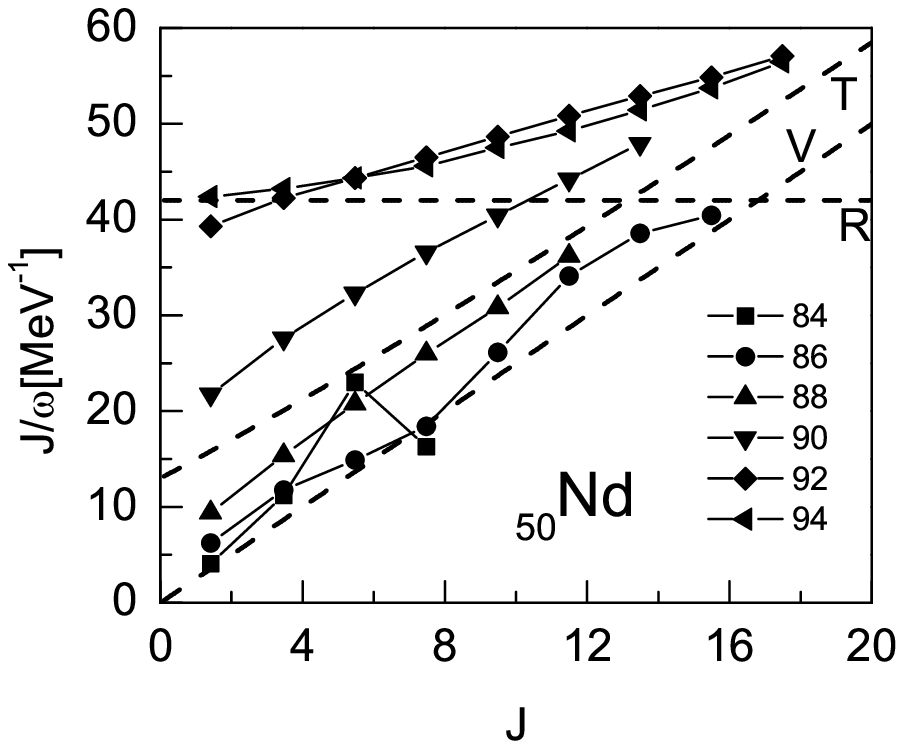}
\caption{Angular momentum $J$ as function of the angular frequency
$\omega$ (upper panel) and kinematic moment
of inertia ${\cal J}$ as function of $J$ (lower panel) 
for the Nd-isotopes. The points correspond to
$\omega(I)=(E(I)-E(I-2))/2$ and $J(I)=\sqrt{I(I-1)}/\omega(I)$,
 where $E(I)$ is the  experimental yrast energy. The 
 dashed lines show the cases discussed in the text.}
\label{f:tidalnd}       
\end{figure}

The quadrupole displacement of the surface with respect to the sphere is 
 $R(t)-R_{sph}=\sum_\mu \alpha_\mu (t) Y_{2\mu}(\vartheta,\varphi)$. The displacement  is expressed by the 
 familiar deformation parameters $\beta, \gamma$ and the vector
$\vec\phi=\vec  n\phi$ (rotation about the axis $\vec n$ by the angle $\phi$), which fixes the orientation of the quadrupole shape.   
\begin{equation}
\alpha_\mu=U^2_{\mu 0}(\vec\phi)\beta\cos{\gamma}+
\left(U^2_{\mu 2}(\vec\phi)+U^2_{\mu -2}(\vec\phi) \right)\frac{\beta\sin{\gamma}}{\sqrt{2}}
\end{equation}
(See \cite{Varshalovich} for the $U^I_{M,M'}$ functions.) 
 The kinetic energy is 
\begin{eqnarray}
2T=D(\dot\beta^2 +\beta^2\dot\gamma^2)+\sum_{i=1}^3\dot\phi^2_i{\cal J}_i\\
{\cal J}_i=4D\beta^2\sin^2(\gamma-i\frac{2\pi}{3})
\end{eqnarray} 
and the potential energy $V(\beta,\gamma)$ depends only on the deformation parameters.
The equations of motion are derived from the Lagrangian $T-V$ in the standard way.
The solution with maximal angular momentum $J$ in z-direction
for given energy (yrast mode) is a wave that
runs over the spherical surface with the constant angular velocity
$\omega$. In the frame of reference rotating with the velocity
$\omega$ the surface has a static deformation. The deformation parameters
$\beta,\gamma$ are the equilibrium values at the minimum of the energy
\begin{eqnarray}\label{Ebeta}
&E(\beta,\gamma)=\frac{J^2}{2{\cal J}(\beta,\gamma)}+V(\beta,\gamma),&\\
&{\cal J}=4D\beta^2\sin^2(\gamma-\frac{2\pi}{3}),&
\end{eqnarray}
where the angular momentum $J={\cal J}\omega$ and $\omega=dE/dJ$.

The location of the droplet surface in spherical 
coordinates is given by
\begin{equation}  \label{tidalclass}
R(t)=R_o[1+\sqrt{2}\beta\sin\gamma\cos(2\phi-2\omega t)Y_{22}(\vartheta,\varphi=0).
\end{equation}  
The inset of Fig. \ref{f:tidalnd} illustrates this mode, which looks like
the tidal waves  on the 
oceans of our planet caused by the gravitational pull of moon and sun.
In allusion, we suggest using this name for
the running waves. Concerning the motion of the surface, it
does not differ from rigid rotation. The difference between
a rotor and a tidal wave is the way angular momentum
is generated. The rigid rotor has a fixed deformation and
a constant moment of inertia ${\cal J}$. The angular momentum 
$J={\cal J}\omega$ is generated
by increasing the angular velocity $\omega$. The tidal
wave runs with a fixed angular velocity $\omega$, while the angular
momentum is gained by increasing the moment of inertia by
 changing $\beta,\gamma$. Of course,
real nuclei lay between these idealized cases. Thus, we suggest the 
following classification: A rotor generates most of its angular momentum $J$ by
increasing  the angular frequency
$\omega$, while the moment of inertia ${\cal J}$ does not change much.
A tidal wave generates most of $J$ by increasing  ${\cal J}$, while
$\omega$ does not change much.   

\begin{figure*}
\begin{minipage}[b]{5.5cm}
\includegraphics[width=\linewidth]{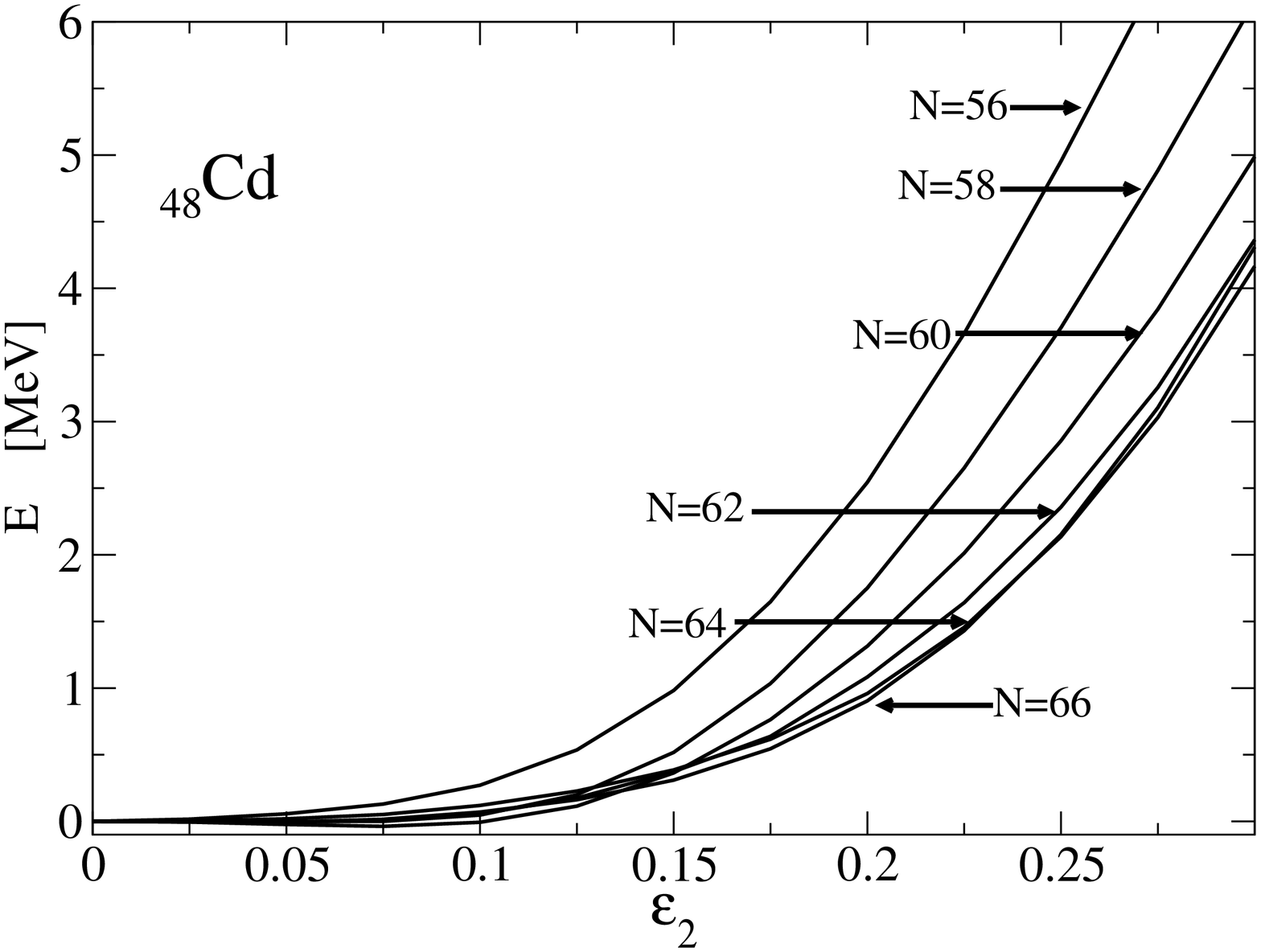}
\end{minipage}
\hfill \mbox{} \hfill
\begin{minipage}[b]{5.5cm}
\includegraphics[width=\linewidth]{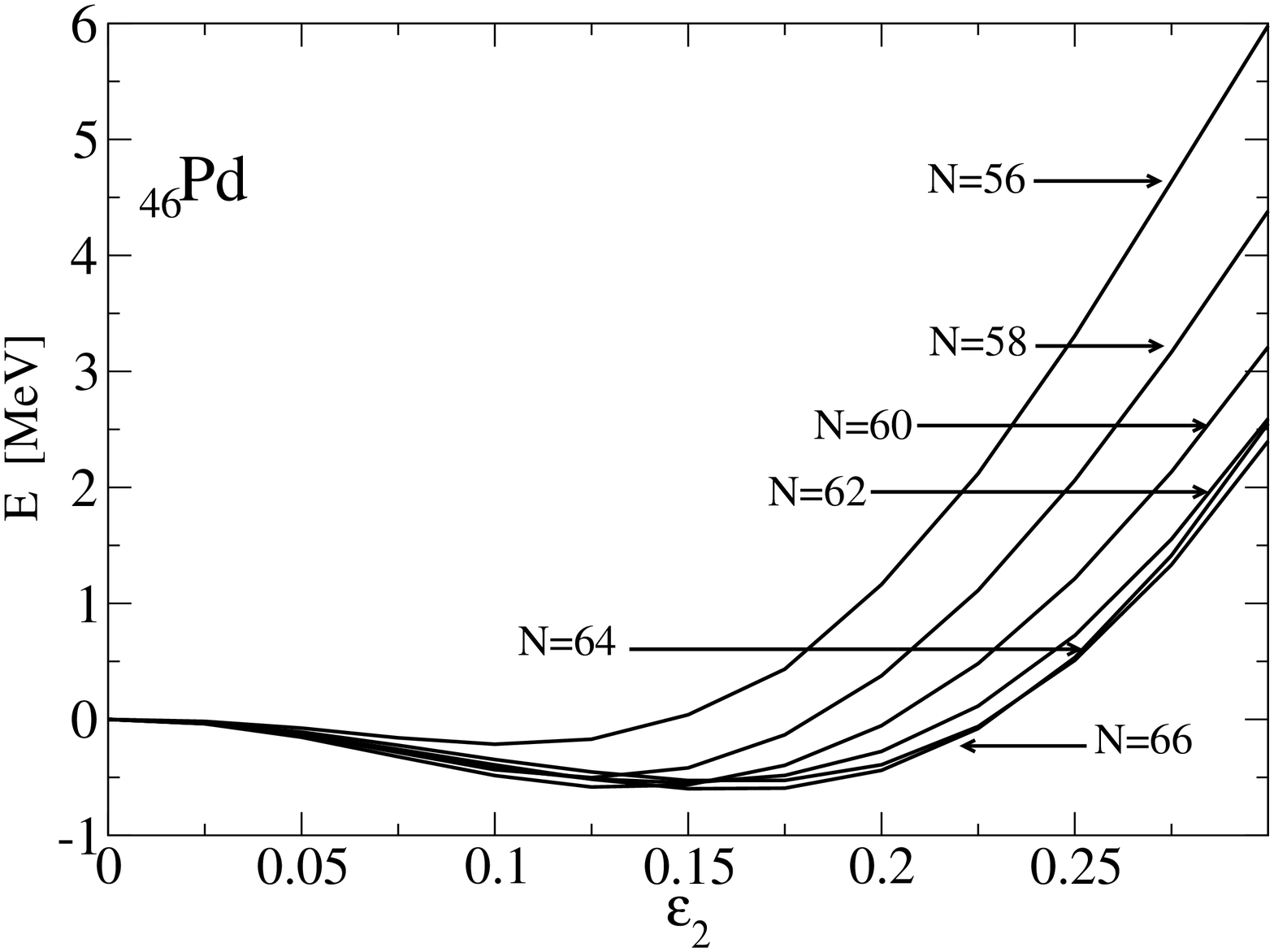}
\end{minipage}
\hfill \mbox{} \hfill
\begin{minipage}[b]{5.5cm}
\includegraphics[width=\linewidth]{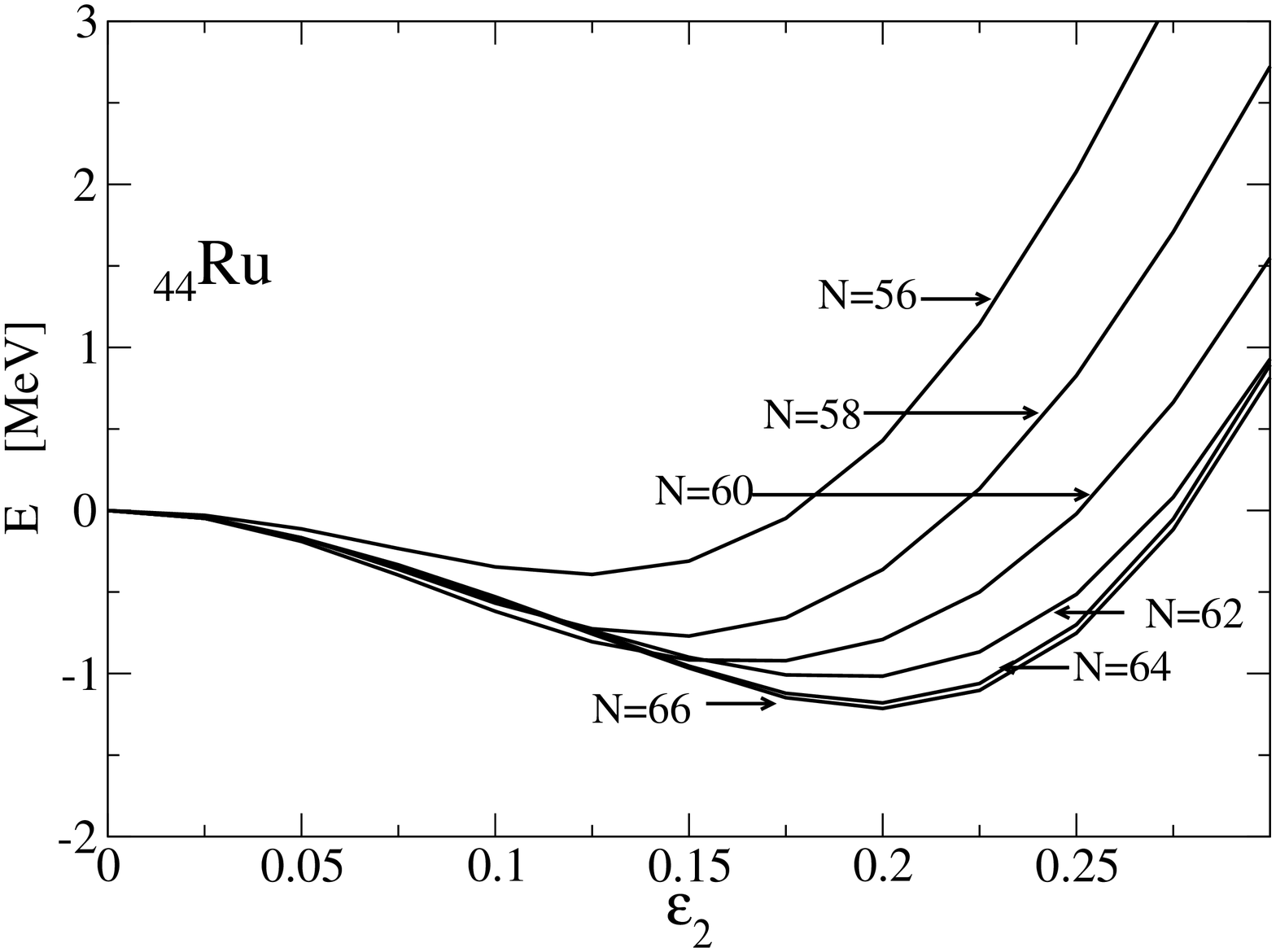}
\end{minipage}
\caption{\label{f:gsenergy}
The ground state energies at $\gamma=5^o$.}
\end{figure*}

Let us discuss some cases in more detail. \\
-Harmonic vibrator (V in  Figs. \ref{f:potential},\ref{f:tidalnd}):\\
\beq\label{Vho}
V=\frac{C}{2}\beta^2,
\eeq
Minimizing the energy one finds
\bea
\gamma=\frac{\pi}{6},~ \beta^2=\frac{J}{2\sqrt{DC}},~ {\cal J}=4D\beta^2=\frac{2J}{\sqrt{DC}},\\
\omega=\frac{J}{{\cal J}}=\frac{1}{2}\sqrt{\frac{C}{D}},~E=\omega J=\Omega\frac{J}{2}=C\beta^2.
\eea
The wave travels with an angular velocity being one half of the oscillator frequency
$\Omega$. The angular momentum is generated by only
increasing the moment of inertia, which is a linear function of $J$. 
This is the ideal
tidal wave. The solution with the same
energy but $J=0$ is $R(t)=R_0(1+\beta \cos(\Omega t)Y_{20}(\vartheta))$, which is the 
familiar oscillation between prolate and oblate shapes.   \\
-Rigid axial rotor (R in Fig. \ref{f:tidalnd}): \\
\beq\label{Vrr}
V=\frac{C\beta_0^2}{4}\left(-\left(\frac{\beta}{\beta_0}\right)^2+\frac{1}{2}\left(\frac{\beta}{\beta_0}\right)^4\right)+V_g\delta(\cos(3\gamma)-1)
\eeq
The potential has a minimum at $\beta_0$, where \mbox{$V\approx C(\beta-\beta_0)^2/2$}.
 Assuming $C$ being infinite large one finds 
\beq
\gamma=-2\pi/3, \beta=\beta_0, {\cal J}_0=3D\beta_0^2=const.
\eeq
-The soft rotor: (R in Fig. \ref{f:potential})\\
With a finite $C$ the minimization gives
\bea
 \beta=\beta_0(1+\frac{2J^2}{{\cal J}_0C\beta_0^2}+O(J^4)),\\
{\cal J}={\cal J}_0(1+\frac{8J^2}{{\cal J}_0C\beta_0^2}+O(J^4)).
\eea
The moment of inertia is a slowly increasing quadratic function of $J$.\\
-Transitional nucleus (T in Figs. \ref{f:potential},\ref{f:tidalnd}):\\
The yrast energies are well approximated by the expression suggested \cite{brentano},
which gives ($J=\sqrt{I(I+1)}$) 
\begin{equation}\label{Ebr}
E(J)=\frac{J^2}{(2(\Theta_0+\Theta_1J)},~{\cal J}=\frac{(\Theta_0+\Theta_1J)^2}{(\Theta_0+\Theta_1J/2)}
\end{equation}
 As expected for a transitional nucleus,
the angular momentum is gained  by increasing both 
$\cal{J}$ and $\omega$. From a vibrational
 perspective, the increase of $\omega$ reflects the anharmonicities
of the motion, from the rotational perspective, the increase
of ${\cal J}$ reflects the softness of the rotor. 
Assuming the energies are derived from minimizing (\ref{Ebeta})
and axial shape ($\gamma=0$) one finds 
 \begin{eqnarray}\label{Vbr}
\frac{\beta(J)}{\beta_0}=\frac{\Theta_0+\Theta_1 J}{\sqrt{B(\Theta_0+\Theta_1 J/2)}},\\
V(J)=\frac{J^2(\Theta_0+\Theta_1 J/2)}{(\Theta_0+\Theta_1 J)^2},
\end{eqnarray}   
which is a parametric form for $V(\beta/\beta_0)$. It is displayed in Fig. \ref{f:potential} as T,
 which has a
form intermediate the vibrator (V) and rotor (R). 
Using $Q_t=Q \beta/\beta_0$ for the  transition quadrupole 
moment of E2-transitions, one may calculate its spin dependence
by means of (\ref{Vbr}). Tab. \ref{t:gd154} shows the case of $^{154}$Gd, which has been 
classified as a transitional point nucleus \cite{Iachello,gd154}. 
The parameters $\Theta_0=21.5$MeV$^{-1}$ and $\Theta_1=1.18$MeV$^{-1}$ place the 
nuclide slightly above the T -line in Fig. \ref{f:tidalnd}. 
The values for $Q_t$ are very similar to the ones
of the X(5) limit given in \cite{gd154}, which is consistent with the   
the shape of the  potential $V(\beta/\beta_0)$. The increase of $Q_{t}$ is larger than 
that of the experimental values $Q_{t,exp}$, which points to degrees of freedom
other than the quadrupole deformation contributing to the 
growth of ${\cal J}(I)$.

The transition between rotation and tidal waves is gradual. 
Let us classify  the yrast mode of  nuclei as an (anharmonic) tidal wave 
when most of the angular
momentum is generated by the increase of the moment of inertia 
and as (soft) rotation when most of the angular momentum is generated by 
the increase of the angular velocity. 
Experimental functions $J(\omega)$ and ${\cal J}(J)$ can be derived from the level energies
by the familiar procedure indicated in the caption of
Fig. \ref{f:tidalnd}, which
shows the Nd-isotopes. For the well deformed $N=94$ isotope, 
the increase of ${\cal J}$ remains moderate compared with ${\cal J}(0)$, and
it grows approximately with $J^2$, which is characteristic for a 
rotor. The frequency $\omega$ starts with a small value and grows steadily. The isotopes with $N=84 - 88$ belong to the region of
 tidal waves.  The frequency $\omega$ starts with a large value and stops growing, while
${\cal J}$ starts with a small value and increases substantially.
As expected, $N=84$ comes 
closest to the spherical  vibrator limit. The   
experimental functions $J(\omega)$  and ${\cal J}(J)$ become increasingly 
irregular when approaching the shell closure at $N=82$, which indicates that the
the motion looses collectivity becoming progressively entangled with the single particle degrees of freedom. The limit of a good vibrator is not realized in nuclei.
The isotop $N=90$ marks the border between a tidal wave and a soft rotor, which is in accordance with the onset of a substantial ground state deformation.

\begin{table}
\caption{Energies $E$ (in MeV) and transition quadrupole moments $Q$
(in eb) for $^{154}$Gd. The subscript {\it pm} is used for the
phenomenological model  given by (\ref{Ebr}) and
(\ref{Vbr}) and {\it e}  for the experiment \cite{ensdf,gd154}. The columns
$\beta/\beta_0$ and $V$ display the deformation potential in parametric form. }
\begin{tabular}{|ccccccc|}
\hline
$I$&$E_{e}$&$E_{pm}$&$\beta/\beta_0$&$V$&$Q_{t,e}$&$Q_{t,pm}$\\
\hline
0&0&0&0&0&-&-\\
2&   0.123&	0.123&	1&	0.11    &6.2	      &	6.2\\
4&0.371&	0.374&	1.07&	0.33	&6.52$\pm$0.06&6.64\\
6&0.717&	0.720&	1.14&   0.62	&6.42$\pm$0.10& 7.05\\
8&1.144&	1.142&	1.20&	0.95    &6.60$\pm$0.16& 7.45\\
10&1.637&       1.623&  1.26&	1.31    & -           &7.84\\
\hline	
\end{tabular}
\label{t:gd154}
\end{table}

The discussion of the liquid drop model for the quadrupole surface modes
suggests that one may calculate the yrast states  in a microscopic
way by finding the selfconsistent  mean field solution in a 
rotating frame of reference, i. e. one may use the cranking model 
for  calculating the tidal wave modes in transitional 
and vibrational nuclei.
In fact, \cite{marshalek} demonstrated that the RPA equations 
for harmonic quadrupole vibrations of nuclei with a spherical 
ground state are a
solution to the selfconsistent cranking model if 
the deformation of the mean field is treated as a small perturbation.
In this paper we solve numerically the cranking problem without the small deformation
approximation, which allows us to describe the yrast states of all 
nuclei in the range between  the harmonic vibrators  and rigid rotors.
Such microscopic approach takes into account
not only the quadrupole shape degrees of freedom, 
as in the above discussed phenomenological model, but also the
the quasiparticle degrees of freedom, which play an important role to
 be discussed below.  Although we only study even-even nuclei in this Letter,
 we want to point out that the method can be applied to
 odd-A and odd-odd nuclei without any problems.

\begin{figure}[t]
\vspace{-1cm}
\includegraphics[width=\linewidth]{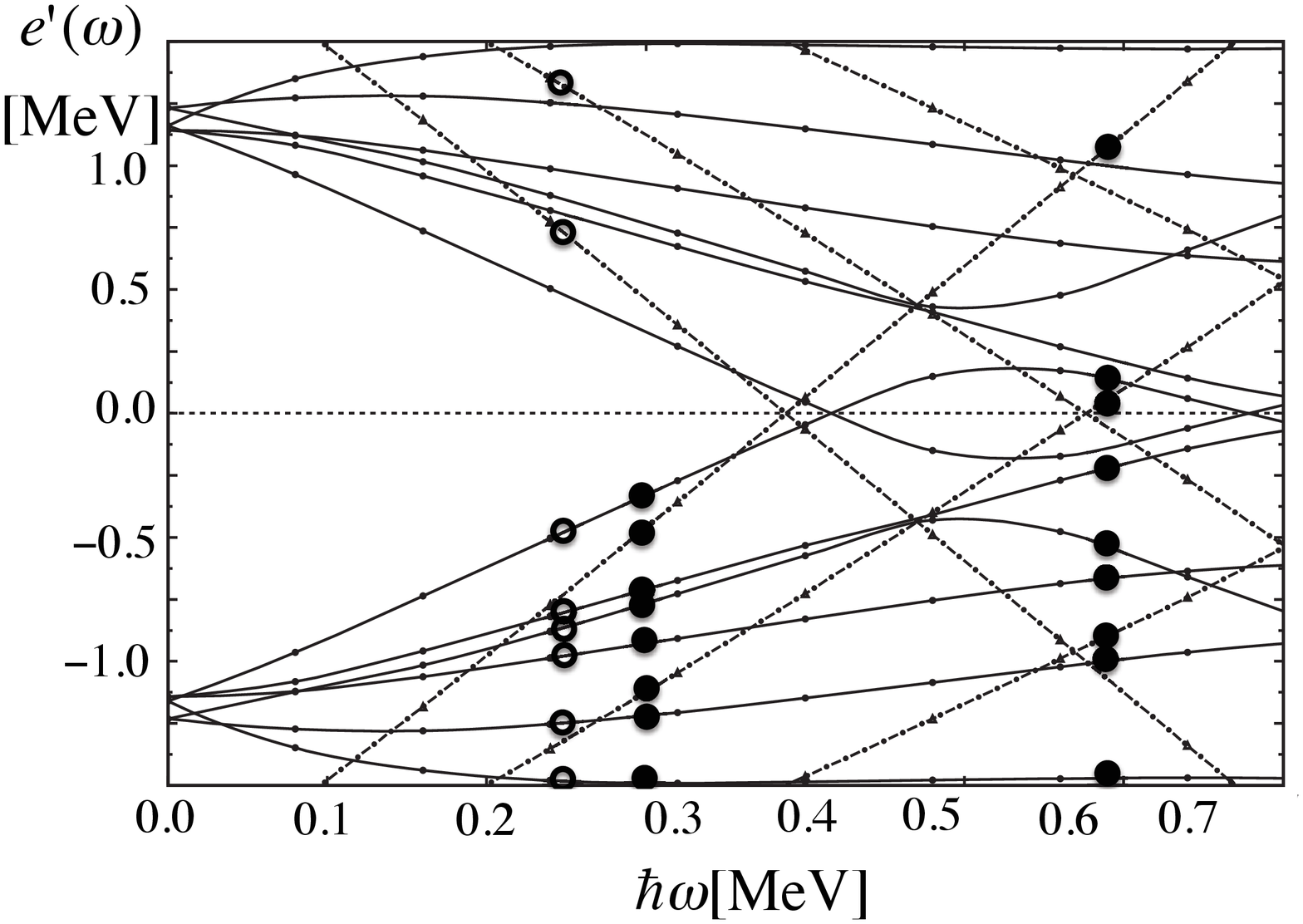}
\vspace*{-1cm}
\caption{The Quasi Neutron Routhians for $N=62$ and \mbox{$\varepsilon_2=0.13$, $\gamma=0$}.
Full lines: positive parity, dash dotted lines: negative parity (h$_{11/2}$). The full circles show the g-configuration,
which corresponds to a tidal wave, the open circles show the s-configuration, which has the character of
antimagnetic rotation. }
\label{f:qneutron}       
\end{figure}

We use
the Shell Correction version of the 
Tilted Axis Cranking model (SCTAC) 
in order to study the tidal waves in the nuclides with 
$Z=$44, 46, 48 and $N=$56, 58, 60, 62, 64, 66. 
The details and parameters of SCTAC are described in \cite{qptac}. 
For the yrast states of even-even nuclei, 
the axis of rotation coincides with a principal axis (x). 
The condition $<J_x>=J=I$ 
is used to fix $\omega$, which is solved numerically for each point on
a grid in the $\varepsilon_2-\gamma$ plane. In solving
it is crucial to employ the diabatic tracing technique described in  \cite{qptac},
which is illustrated in Fig. \ref{f:qneutron}. The ground state (g-)
configuration indicated by the full dots is followed through
crossings of the h$_{11/2}$ routhians. 
The avoided crossings between the positve parity routhians 
around $\hbar \omega=0.53$ MeV are taken into account adiabatically:
The occupation is kept on the lower of the two levels. 
The yrast states with high spin correspond to the   s-configuration
indicated by the open circles  

\begin{figure}[tb]
\includegraphics[width=\linewidth]{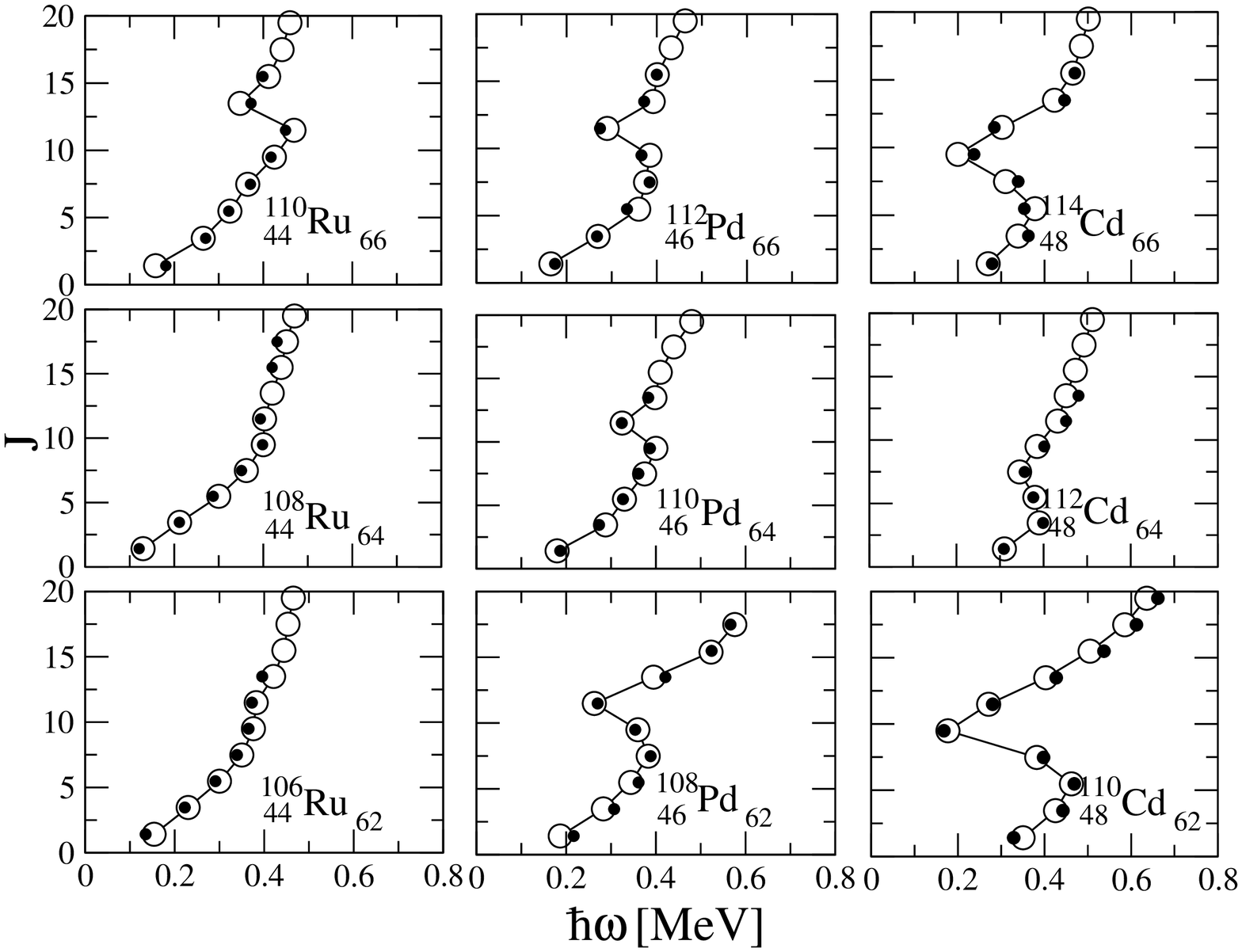}

\vspace*{0.9cm}

\includegraphics[width=\linewidth]{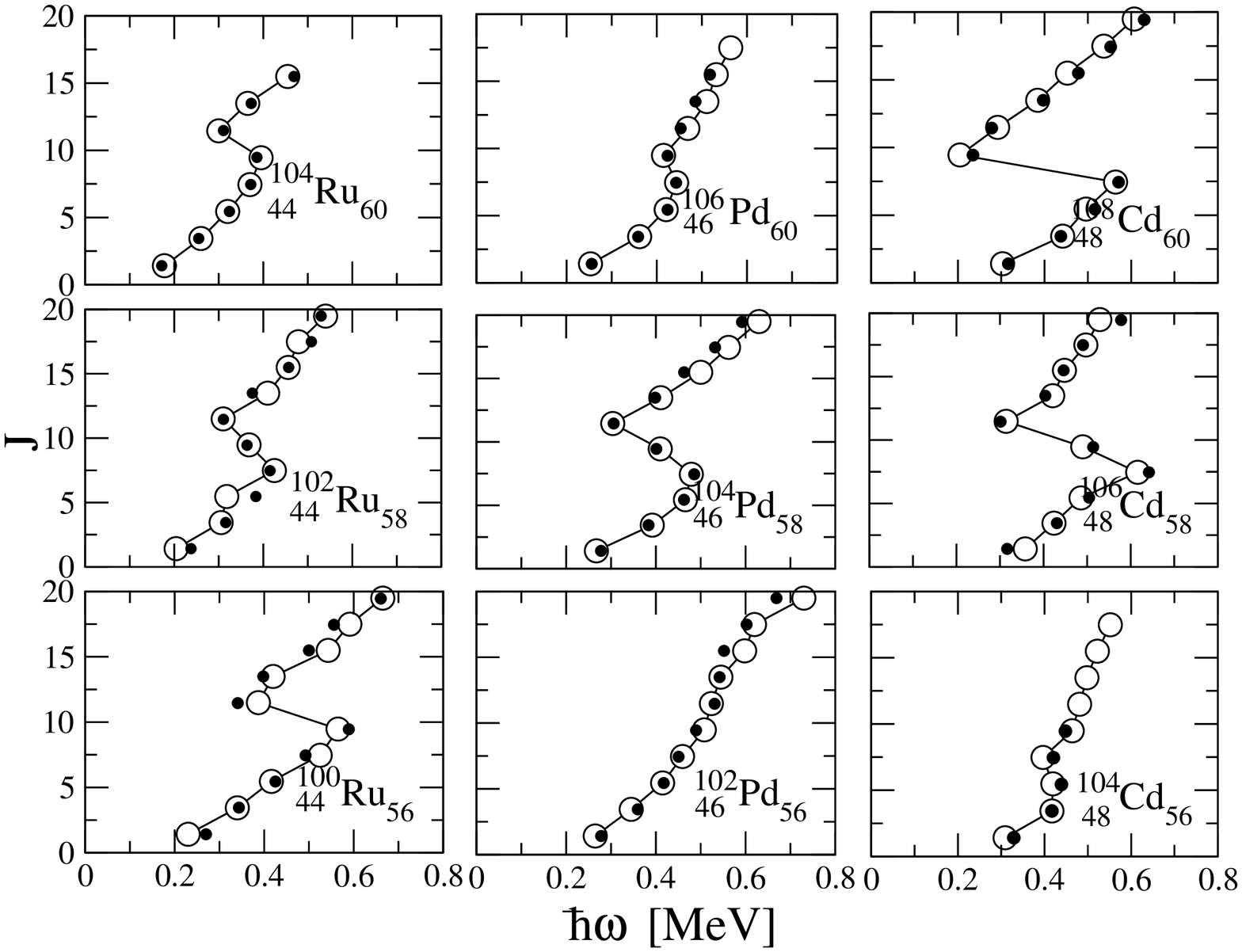}
\caption{Angular momentum $J$ as function of the angular frequency
$\omega$ in the mass 110 region. The quantities $J$ and $\omega$
are obtained as in Fig. \ref{f:tidalnd} from the calculated
or experimental yrast energies $E(I)$. 
 Full symbols show the experiment, open the calculations.
 Data from \cite{ensdf}.}
\label{f:tidaliom}       
\end{figure}

\begin{figure}[t]
\includegraphics[width=\linewidth]{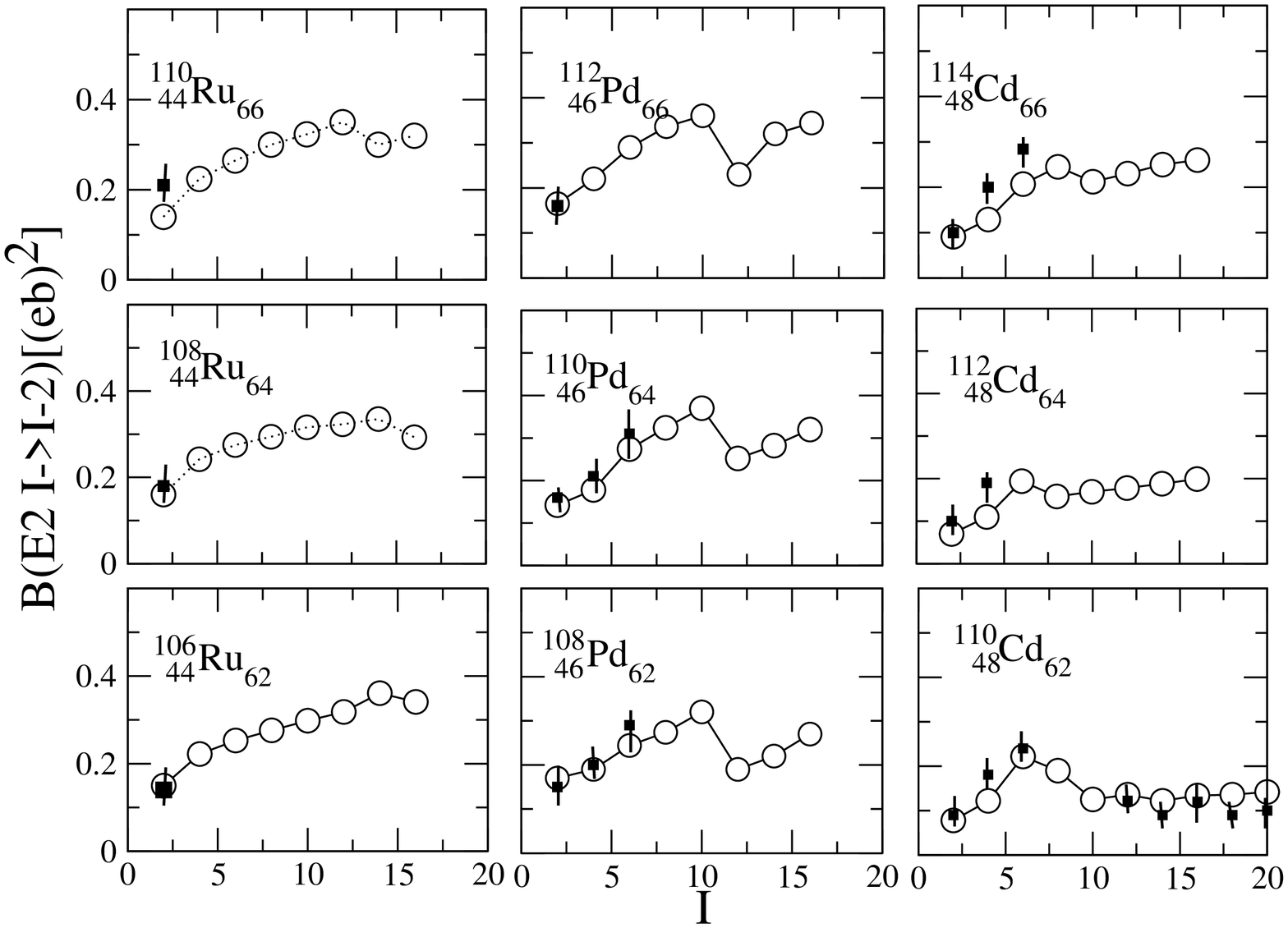}

\vspace*{1.3cm}

\includegraphics[width=\linewidth]{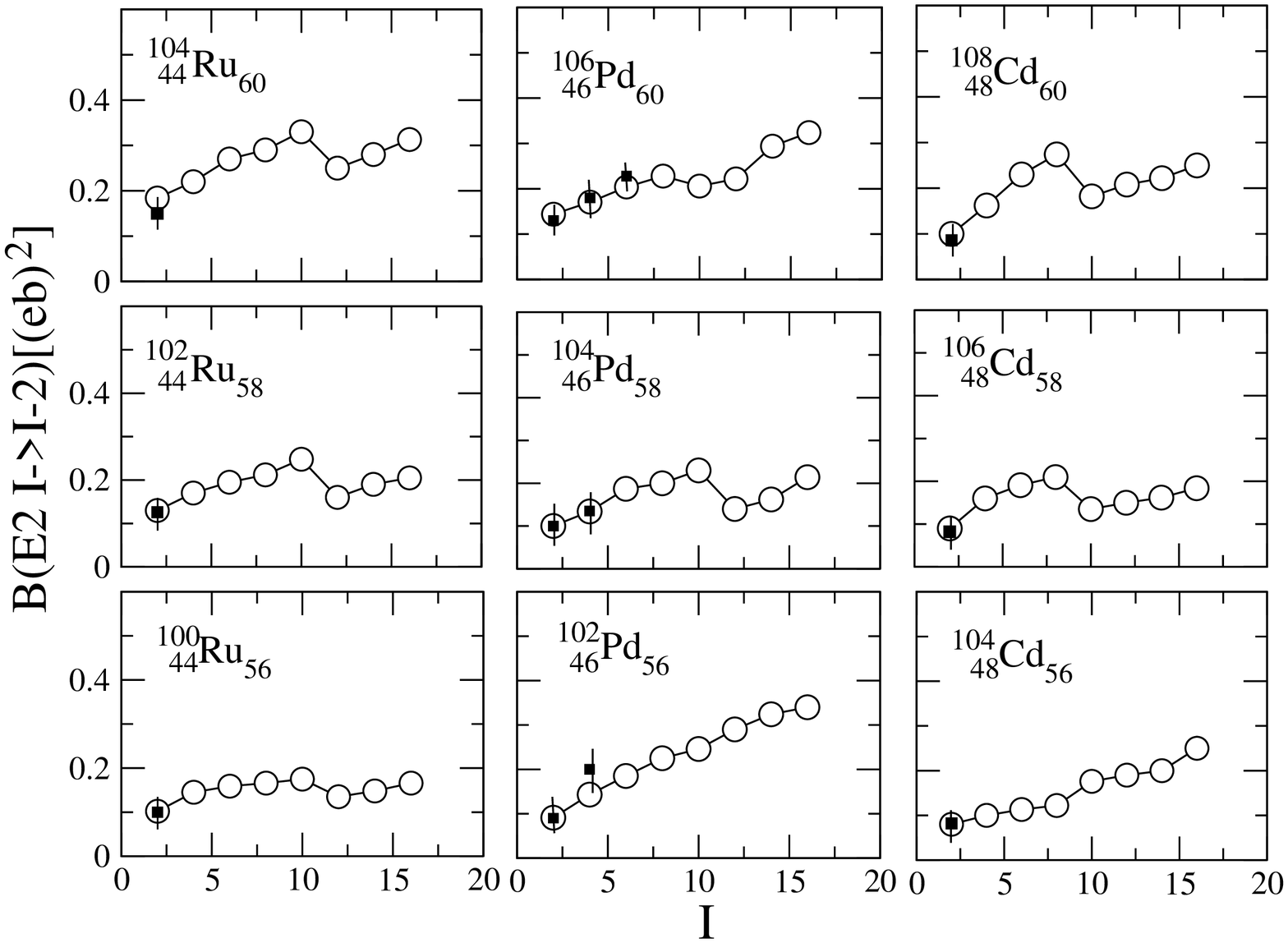}

\vspace*{0.15cm}

\caption{The $B(E2, I->I-2)$ values in the mass 110 region.
Full symbols show the experiment, open the calculations. Data from
\cite{raman,ensdf,108pde2,102pde2,104pde2,108cde2,110cde2,RMP}.}
\label{f:tidalbe2}       
\end{figure}

The interpolated function $E(\varepsilon_2,\gamma,J)$ 
is minimized with respect to deformation parameters for 
fixed angular momentum. The final 
value of $\omega(J)$ for the equilibrium deformation is
found by interpolation between the values at the grid points.
It is shown as the theoretical points in the Figure \ref{f:tidaliom}.
It is noted, that the standard technique of finding the selfconsistent
solutions by iteration at fixed $\omega$ becomes problematic
for tidal waves close to the harmonic limit because the total Routhian is 
nearly deformation independent.           
The pairing parameters  are kept fixed to $\Delta_n=\Delta_p=1.2$ MeV, 
 and the chemical
potentials are $\lambda$ adjusted to the particle numbers at $\omega=0$. 
The $B(E2,I->I-2)$ values are calculated for $J=I$ as
described in \cite{qptac} for $\theta=90^o$ and multiplied
by $I(I-1)/(I-3/2)(I-1/2)$. The factor corrects for the low-spin
deviation of the $B(E2)$ values from the high-spin limit used 
in the TAC code. It represents the Clebsh-Gordan coefficient
appearing for axial $K=0$ bands  \cite{BMII}, which is justified
by the fact that for most nuclei $\gamma <10^o$. 
We slightly correct
\beq
 J(\omega)=J_{SCTAC}(\omega)+ 100MeV^{-1} \varepsilon_2^2\sin^2(\gamma-\pi/3),
\eeq
which adds less than 0.5 $\hbar$ to the angular momentum. About half of 
it takes into account the coupling between the oscillator
shells  and another half is expected
to come from quadrupole pairing, both being neglected in SCTAC. 
The changes due to corrections are smaller than the radius of the
open circles in Figs. \ref{f:tidaliom} and \ref{f:tidalbe2}.

The ground state energies are shown  in Fig. \ref{f:gsenergy}. They are 
calculated using the modified oscillator potential with the parameters
given in Ref. \cite{qptac}. 
The tendency to deformation increases with the numbers of 
valence proton holes and valence neutron particles in the $Z=N=50$ 
shell. The $Z=48$ isotopes are spherical, becoming softer 
with increasing $N$.   
The $Z=44$ isotopes are slightly deformed, very soft, becoming less
soft with $N$. The $Z=46$ isotopes are intermediate. 
Figs. \ref{f:tidaliom} and \ref{f:tidalbe2} compare
  the calculated values of $J(\omega)$ and $B(E2, I->I-2)$ with experiment.
The detailed agreement was achieved by adding to ground state deformation energies
the small adjustment $k \varepsilon_2^2$ with 
$k=$8.5, 7.9, 8.5, 9.6, 11.9, 14.5 MeV for $N$= 56, 58, 60, 62, 64, 66, respectively, which makes the potentials 
slightly stiffer.
The correction term is
substantially smaller  than the differences between deformation energies 
calculated from the various mean field theories used at present.
Tab. \ref{t:cd110} compares calculations with and  without the correction term.
\begin{table}
\caption{Rotational frequencies in$ ^{110}$Cd. The subscript o is used for 
the calculation without the correction
of the ground state energies, the subscript c for  the one with the correction, and the
subscript e for the
experiment. The calculated energy
$E(0^+_2)$=3.51 MeV, its experimental value 1.47 MeV. All energies are in MeV. Data from \cite{ensdf}}
\vspace*{0.3cm}
\begin{tabular}{|cccccc|}
\hline
$I$&$\hbar \omega_o(0^+_1)$&$\hbar \omega_c(0^+_1)$&$\hbar \omega_{e}(0^+_1)$&$\hbar \omega_{o}(0^+_2)$&$\hbar \omega_{e}(0^+_2)$\\
\hline
2&   0.336&	0.340&0.328	&0.187&	0.155   \\
4&  0.381&	0.410&0.442	&0.260&	0.233	\\
6&  0.400&	0.430&0.468	&0.295&   0.313	\\
8&  0.381&	0.390&0.398	&           &	  \\
10&0.194&        0.190&0.168     &           &                \\
12&0.283&        0.280& 0.280    &           &                  \\
\hline
\end{tabular}
\label{t:cd110}	
\end{table}   

 As seen in Fig. \ref{f:tidalbe2}, for low-spin ($I\leq 6$) the 
tidal waves in Cd isotopes  show the linear relation 
$B(E2)\propto J$ that is expected for vibrational nuclei. However, the
energies deviate strongly from the vibrator limit. For Ru and Pd,
the  $B(E2)$ values start at larger values and increase less with $I$, such that they
do not extrapolate to the origin. This reflects the transition to 
more rotational behavior, which becomes stronger with increasing $N$.
The low-spin properties reflect the ground state deformation 
energies in Fig. \ref{f:gsenergy}.

At high spin, there is an irregularity in the $J(\omega)$ curves, which
is caused by the rotational  
alignment of a pair of $h_{11/2}$ neutrons (change from the g-configuration to the s- configuration in Fig. \ref{f:qneutron}). 
It is well known that
the sharpness of the alignment depends
sensitively on the position of the neutron chemical potential $\lambda_n$ 
relative to the  $h_{11/2}$ levels \cite{Hamamoto}, which causes the familiar
oscillations between rapid alignment (back bending) and more gradual one (vertical section)
 with increasing $N$. The $h_{11/2}$ levels  move with the changing deformation, and
differences of the deformation are the reason
why $J(\omega)$ sharply back bends in  $Z=48,~N=62$ while it gradually grows
in $Z=44,~N=62$. Also normal parity neutron orbitals contribute to the $J(\omega)$
in a non-linear way, in particular at the avoided crossings  between them 
(around $\hbar \omega = $ 0.53 MeV in Fig. \ref{f:qneutron}).
 For example, they are responsible that in $^{110}$Cd
 the  point $I=8$ has a lower $\omega$ than the point $I=6$. 
 The same is observed in $^{108}$Cd \cite{108cde2} (not shown). The large frequency
encountered at small deformation makes also other orbitals than the high-j intruders react
in a non linear way to the inertial forces.  For many nuclides 
the quasiparticle degrees cannot be treated in a perturbative way for $I \geq 6$, which means
the separation of the collective quadrupole degrees of freedom becomes
problematic. For the nuclei with the smallest numbers of valence particles
and holes this entanglement of collective and single particle degrees appears
already for $I=4$, which is reflected by the irregular curves $J(\omega)$ for
$Z=48,~N=56$, and $Z=60,~N=84$ in Figs. \ref{f:tidaliom}, and \ref{f:tidalnd}, respectively. 

As seen in Fig. \ref{f:tidalbe2},
the aligned $h_{11/2}$ quasi neutrons
 stabilize the deformation at a {\it smaller}
 value than reached before, which is reflected by the decrease 
of the $B(E2)$ values. The deformation increases slower than before the
alignment, which means that the motion becomes more rotational.
In fact, it takes the character of Antimagnetic Rotation, discussed in \cite{RMP}.  
In some cases, the alignment process is extended over several
units of angular momentum, keeping $\omega$ nearly constant.
This means the simple picture of tidal waves, as running at nearly
 constant angular velocity and gaining angular momentum by 
deforming does not account for the 
full complexity of real nuclei. Other degrees of freedom,
in particular the single particle ones, may also 
provide angular momentum such that the rotational frequency remains
constant.         

 $^{102}$Pd is a special case.  The g- and s-configurations do not mix and are both
 observed up to $I=20$. Figs. \ref{f:tidaliom} and \ref{f:tidalbe2} show only the 
 g-sequence, which represents a tidal wave that gains angular momentum 
 predominantly by increasing the deformation, as discussed above for the
 droplet model. Its amplitude at $I=16$ corresponds to eight stacked phonons! 
 Only for   $I >16$ the alignment of the positive parity neutrons 
 (avoided crossing at \mbox{$\hbar \omega \approx 0.53$ MeV} in Fig. \ref{f:qneutron})
 becomes a comparable source of angular momentum.

Of course, our cranking approach also allows us to calculate rotational
bands built on quasiparticle excitations. As an example,
Tab. \ref{t:cd110} contains the lowest two-quasiproton excitation across
the $Z=50$ shell gap. The excited quasiprotons drive the nucleus to a
  substantial deformation, which, in contrast to one of the yrast tidal wave
  states,  weakly increases with spin. The 0$^+_2$ "intruder state" 
  has been assigned to this configuration, whose deformed shape  coexists
  with the near spherical ground state shape \cite{cd110intruder}.
  The energies  of the rotational band are well reproduced by the calculation,
  which however gives a too high energy of the $0^+_2$ state.


In conclusion, the yrast states of vibrational
and transitional nuclei can be understood as tidal waves that run over the nuclear 
surface, which have a static deformation, the wave amplitude, in the frame of 
reference co-rotating
with the wave. In contrast to rotors,  most of the angular momentum is 
gained by an increase of the deformation  while the angular velocity
increases only weakly. Since tidal waves have a static deformation in the rotating frame,
the rotating mean field approximation (Cranking model) provides a microscopic
description. The yrast states 
of  ``spherical'' or weakly deformed nuclei with $44\leq Z\leq 48,~56\leq N\leq66$ were calculated up to
spin 16. These microscopic calculations reproduce the energies and 
electromagnetic $E2$ transition probabilities in detail.
 The structure of the yrast states turns out to be complex. The shape
degrees of freedom and the single particle degrees of freedom are intimately
 interwoven. The non-linear response to the inertial forces
of the individual orbitals at the Fermi surface determines the
way the angular momentum is generated. This lack of separation 
of the collective motion from the single particle becomes increasingly important 
with decreasing number of valence nucleons. It can set as low as 
$I=6$ already.
Purely collective models like IBA \cite{IBM} or the Dynamical Collective 
Model \cite{DCM}
miss this explicit coupling between single particle and collective freedoms.

Supported by the DoE Grant DE-FG02-95ER4093.

\end{document}